\documentstyle[12pt]{article}
\def\beq{\begin{equation}}
\def\eeq{\end{equation}}

\def\ap#1#2#3 {Ann. Phys. (NY) {\bf#1} (19#2) #3}
\def\apj#1#2#3 {Astrophys. J. {\bf#1} (19#2) #3}
\def\apjl#1#2#3 {Astrophys. J. Lett. {\bf#1} (19#2) #3}
\def\app#1#2#3 {Acta. Phys. Pol. {\bf#1} (19#2) #3}
\def\ar#1#2#3 {Ann. Rev. Nucl. Part. Sci. {\bf#1} (19#2) #3}
\def\cpc#1#2#3 {Computer Phys. Comm. {\bf#1} (19#2) #3}
\def\err#1#2#3 {{\it Erratum} {\bf#1} (19#2) #3}
\def\ib#1#2#3 {{\it ibid.} {\bf#1} (19#2) #3}
\def\jmp#1#2#3 {J. Math. Phys. {\bf#1} (19#2) #3}
\def\ijmp#1#2#3 {Int. J. Mod. Phys. {\bf#1} (19#2) #3}
\def\jetp#1#2#3 {JETP Lett. {\bf#1} (19#2) #3}
\def\jpg#1#2#3 {J. Phys. G. {\bf#1} (19#2) #3}
\def\mpl#1#2#3 {Mod. Phys. Lett. {\bf#1} (19#2) #3}
\def\nat#1#2#3 {Nature (London) {\bf#1} (19#2) #3}
\def\nc#1#2#3 {Nuovo Cim. {\bf#1} (19#2) #3}
\def\nim#1#2#3 {Nucl. Instr. Meth. {\bf#1} (19#2) #3}
\def\np#1#2#3 {Nucl. Phys. {\bf#1} (19#2) #3}
\def\pcps#1#2#3 {Proc. Cam. Phil. Soc. {\bf#1} (#2) #3}
\def\pl#1#2#3 {Phys. Lett. {\bf#1} (19#2) #3}
\def\prep#1#2#3 {Phys. Rep. {\bf#1} (19#2) #3}
\def\prev#1#2#3 {Phys. Rev. {\bf#1} (19#2) #3}
\def\prl#1#2#3 {Phys. Rev. Lett. {\bf#1} (19#2) #3}
\def\prs#1#2#3 {Proc. Roy. Soc. {\bf#1} (19#2) #3}
\def\ptp#1#2#3 {Prog. Th. Phys. {\bf#1} (19#2) #3}
\def\ps#1#2#3 {Physica Scripta {\bf#1} (19#2) #3}
\def\rmp#1#2#3 {Rev. Mod. Phys. {\bf#1} (19#2) #3}
\def\rpp#1#2#3 {Rep. Prog. Phys. {\bf#1} (19#2) #3}
\def\sjnp#1#2#3 {Sov. J. Nucl. Phys. {\bf#1} (19#2) #3}
\def\spj#1#2#3 {Sov. Phys. JEPT {\bf#1} (19#2) #3}
\def\spu#1#2#3 {Sov. Phys.-Usp. {\bf#1} (19#2) #3}
\def\zp#1#2#3 {Zeit. Phys. {\bf#1} (19#2) #3}
\begin{document}
\begin{titlepage}
\begin{center}
{\Large \bf Theoretical Physics Institute \\
University of Minnesota \\}  \end{center}
\vspace{0.3in}
\begin{flushright}
TPI-MINN-94/38-T \\
UMN-TH-1320-94 \\
November 1994
\end{flushright}
\vspace{0.4in}
\begin{center}
{\Large \bf Moments of lepton spectrum in $B$ decays and the $m_b - m_c$
quark mass difference. \\}
\vspace{0.2in}
{\bf M.B. Voloshin  \\ }
Theoretical Physics
Institute, University of Minnesota \\ Minneapolis, MN 55455 \\ and \\
Institute of Theoretical and Experimental Physics  \\
                         Moscow, 117259 \\
\vspace{0.3in}
{\bf   Abstract  \\ }
\end{center}
It is argued that the quark mass difference $m_b-m_c$ can be extracted with
a high accuracy from experimental data on ratia of moments of lepton energy
spectrum in semileptonic decays of $B$ mesons. Theoretical expressions for
the moments are presented, which include perturbative as well as
non-perturbative corrections.

\end{titlepage}

\section{Introduction}
Masses of quarks as well as weak mixing angles are fundamental input
parameters in the Standard Model. Thus it is important to know them with
maximal possible accuracy. Moreover, the values of the masses of the
$b$ and $c$ quarks are correlated with the determination of the mixing
parameter $|V_{bc}|$ from the data on inclusive semileptonic
$B$ meson decay rates. Therefore an independent understanding of $m_b$ and
$m_c$ is necessary for a precision determination of $|V_{bc}|$. There is a
vast literature on extracting the values of $m_c$ and $m_b$ from the data
on charmonium, $\Upsilon$ resonances, charmed and $B$ hadrons. However
there is hardly a compelling argument in the literature to invalidate the
original evaluations from the QCD sum rules of the `on shell' quark masses:
$m_c = 1.35 \pm 0.05 \, GeV \,^{\cite{csr1,csr2,csr3}}$ and $m_b=4.80 \pm
0.03 \, GeV \, ^{\cite{bsr}}$, which can still serve as `reference values'
in discussion of dynamics of heavy hadrons. It is the purpose of this paper
to point out that the accuracy of determining the difference between the
quark masses $m_b-m_c$ can possibly be significantly improved by
considering the ratia of the moments
\beq
M_n=\int E_l^n {{d \Gamma} \over {d E_l} } \, d E_l
\label{mn}
\eeq
of the lepton energy spectrum $d \Gamma/ d E_l$ in semileptonic $B$ decays.
While theoretical expressions for the moments are sensitive to combinations
of $m_b$ and $m_c$ (and also to $|V_{bc}|$) ,
their ratia for few first moments are to a high accuracy sensitive only to
the mass difference $m_b - m_c$. Also it is natural to expect that
experimentally the ratia of the moments are determined with better
systematic accuracy, since the absolute normalization of the event rate
cancels out in the ratia.

On the theoretical side the ratia of the moments have the advantage of weak
and controlled dependence on the infrared dynamics in QCD both
perturbatively and non-perturbatively. For first few moments the
perturbative corrections are expressed through $\alpha_s(m_b)$ and the
non-perturbative ones are suppressed by $m_b^{-2}$ and can be found by
the OPE technique$^{\cite{vs,bvu,bsvu,bksv}}$ in terms of the quantities
$\mu_\pi^2/m_b^2$ and $\mu_g^2/m_b^2$ with
\beq
\mu_\pi^2 = \langle B |({\overline b} \mbox{\boldmath $\pi$}^2
b)| B \rangle ~~~~~~ {\rm
and} ~~~~~~ \mu_g^2= \langle B |\left({\overline b} \,
(\mbox{\boldmath $\sigma \cdot B$}) \, b \right)| B \rangle~,
\label{mupg}
\eeq
where {\bf $B$} is the chromo - magnetic field operator and
{\boldmath $\pi\,$}$= {\bf p} - {\bf A}$ is the covariant momentum
operator for the heavy quark. The spin-dependent chromo - magnetic energy
$\mu_g^2$ is related to the mass splitting of $B^*$ and $B$ mesons:
$\mu_g^2={3 \over 4} \, (M_{B^*}^2-M_B^2) \approx 0.36 \, GeV^2$, while
for the kinetic energy $\mu_\pi^2$ only a lower bound exists$^{\cite{v}}$:
$\mu_\pi^2 \ge \mu_g^2$, which follows from non-negativity of the operator
$(\mbox{\boldmath $\sigma \cdot \pi$})^2=\mbox{\boldmath $\pi$}^2 -
\mbox{\boldmath $\sigma \cdot B$}$ and an estimate$^{\cite{bb}}$ $\mu_\pi^2
\approx
0.5 \pm 0.1 \, GeV^2$ from QCD sum rules.

It should be also noticed that the difference $m_b - m_c$, unlike each of
the masses, is less sensitive to the infrared behavior in QCD and is a well
defined quantity in QCD in the limit where both masses are heavy as
compared to $\Lambda_{QCD}$. Indeed, because of confinement there is no
real `mass shell' for a quark. Therefore its mass can only be determined
off shell and then extrapolated to a would-be on-shell value. For a heavy
quark its mass can be found at a virtuality scale $\mu$ (i.e. at
$m^2-p^2 \approx 2\,m\, \mu$) such that on one hand $\mu \gg
\Lambda_{QCD}$, which justifies a short-distance treatment, and on the
other hand $\mu \ll m$. The latter condition ensures that the evolution
of $m(\mu)$ towards the would-be mass shell does not depend on $m$ in the
leading order in $1/m$. In particular, in the leading log approximation this
evolution is described by the RG equation$^{\cite{v2,bsvu2}}$
\beq
{{d \, m} \over {d \mu}}= - c \, \alpha_s(\mu)~,
\label{rgm}
\eeq
where the constant $c$ depends on the specific definition of the off-shell
mass.
The infrared singularity of $\alpha_s$ (infrared renormalon) prevents from
integrating an equation like (\ref{rgm}) down to $\mu=0$ and thus really
extrapolating the mass to the mass shell. However one can integrate the
equation (\ref{rgm}) in any finite order in $\alpha_s$ and thus define the
`on-shell' mass of a heavy quark to a finite order. In this sense the
`on-shell' masses $m_b$ and $m_c$ quoted above are the result of such
extrapolation in the first order and are thus appropriate for using in
other calculations in the first order in $\alpha_s$.
Naturally this definition of quark mass
changes with the order in $\alpha_s$. However, increasing the order in
$\alpha_s$ does not converge at a certain value of $m$ because of the
factorial divergence of the series in $\alpha_s$ caused by the infrared
renormalon. A minimal residual error in the `on shell' mass in this
procedure is of the order of $\Lambda_{QCD}\,^{\cite{bsvu2}}$. On the other
hand this uncertainty in a heavy quark mass does not depend on $m$ in the
limit of large $m$. Thus this uncertainty cancels in the difference of
masses of two heavy quarks. As to the pre-asymptotic in the heavy quark
mass limit corrections to the evolution equation (\ref{rgm}), their
contribution to the residual uncertainty is of the order of
$\Lambda^3_{QCD}/m^2$, which is quite small even for the charmed quark.

The difference $m_b - m_c$ can be estimated from the experimental
values of the masses of $D$ and $B$ mesons:
\beq
M_B-M_D = m_b - m_c + {{\mu_\pi^2-\mu_g^2} \over 2\,m_b}-
{{\mu_\pi^2-\mu_g^2} \over 2\,m_c}+o(m_c^{-2},\,m_b^{-2})
\label{diff}
\eeq
Neglecting the terms, smaller than $m_c^{-2}$ or $m_b^{-2}$, and taking
into account the inequality $\mu_\pi^2 \ge \mu_g^2$ one finds a lower bound
for the difference of the quark masses:
\beq
m_b-m_c \ge M_B-M_D = 3.41\, GeV~.
\label{lb}
\eeq
Varying $\mu_\pi^2$ in the range from $\mu_\pi^2 = \mu_g^2 \approx 0.36\,
GeV^2$ up
to $\mu_\pi^2 = 0.6 \, GeV^2$ one finds $m_b-m_c = 3.44 \pm 0.03 \, GeV$,
which is perfectly compatible with the quoted above estimates of each
of the quark masses from QCD sum rules.

Naturally, an independent measurement of this quark mass difference with a
comparable or better accuracy would provide an additional consistency check
for the heavy quark theory and, possibly, would enable a better quantitative
understanding of the parameter $\mu_\pi^2$. As is discussed in the rest of
this paper, a measurement of the ratia of few first moments (\ref{mn})
provides an excellent opportunity to independently determine the difference
$m_b-m_c$.

\section{Moments of the lepton spectrum}

In the simplest approximation, where the QCD effects are neglected
altogether the spectrum of charged lepton $l$ in the decay $b \to c \, l
\, \nu$ is given by the well-known muon decay formula
\beq
{{d \, \Gamma} \over dx}= \Gamma_0 \,w_0(x,\,\mu)
\label{g0}
\eeq
with $\Gamma_0=G_F^2\, |V_{cb}|^2 \,m_b^5/(192\, \pi^3)$ and
\beq
w_0(x,\, \mu)= {{2\,x^2 \, (1-\mu^2-x)^2} \over {(1-x)^3}} \left [ (1-
x)\,(3-2x)+ \mu^2 \, (3-x) \right ]~,
\label{w0}
\eeq
where $\mu=m_c/m_b$ and $x=2\, E_l /m_b$, so that the physical range of
$x$ goes from $x_m=0$ to $x_M=1-\mu^2$.

With the first perturbative QCD correction and the first
non-perturbative corrections, proportional to $\mu_\pi^2/m^2$ and
$\mu_g^2/m^2$, taken into account the formula for the differential decay
rate can be written as
\beq
{1 \over \Gamma_0}\,
{{d \, \Gamma} \over dx}= w_0(x,\,\mu)-{2 \over 3} {\alpha_s \over
\pi}\,
w_1 (x,\, \mu) + {\mu_\pi^2 \over m_b^2}\,w_\pi(x,\mu)+{\mu_g^2 \over
m_b^2}\, w_g(x,\mu)
\label{g1}
\eeq
where the explicit expression for the perturbative correction
function $w_1 (x,\, \mu)$ is extremely lengthy and can be found in
the original papers${\cite{jk,cj}}$ (see also \cite{corbo,hp})\footnote{
I am thankful to M. Je\.zabek, for pointing out to me the papers \cite{jk}
and \cite{cj}, where the calculation of the function $w_1(x,\, \mu)$ has
been finalized, and for sending me his and A. Czarnecki's FORTRAN code
for numerical calculation of this function.}.
The non-perturbative  correction functions
$w_\pi(x,\,\mu)$ and $w_g(x,\,\mu)$ are given by$^{\cite{bsvu,bksv}}$
\begin{eqnarray}
 w_\pi (x,\, \mu) =
{{2\,x^3}\over
   {3\,{{\left( 1 - x \right) }^5}}} &
\left( -5 - 15\,{\mu^4} + 20\,{\mu^6} + 25\,x + 21\,{\mu^4}\,x -
       10\,{\mu^6}\,x - 50\,{x^2} -
\right.
\nonumber \\
&\left. 6\,{\mu^4}\,{x^2} + 2\,{\mu^6}\,{x^2} +
       50\,{x^3} - 25\,{x^4} + 5\,{x^5} \right)~,
\label{wpi}
\end{eqnarray}
\begin{eqnarray}
w_g(x,\,\mu)=
{{2\,x^2\,\left( 1 - {\mu^2} - x \right)}\over
   {3\,{{\left( 1 - x \right) }^4}}}
& \left( 6 - 12\,{\mu^2} - 30\,{\mu^4} - 13\,x + 23\,{\mu^2}\,x +
       20\,{\mu^4}\,x + 3\,{x^2} - \right.
\nonumber \\
& \left.16\,{\mu^2}\,{x^2} - 5\,{\mu^4}\,{x^2} +
       9\,{x^3} + 5\,{\mu^2}\,{x^3} - 5\,{x^4} \right)~.
\label{wg}
\end{eqnarray}

The relative correction
$w_1 (x,\, \mu)/w_0 (x,\, \mu)$ has a logarithmic singularity at the upper
endpoint $x=x_M$ of the spectrum$^{\cite{corbo}}$, which is a usual
consequence of the Sudakov form-factor$^{\cite{fjmw}}$. The relative
non-perturbative corrections are still more singular:  the ratio \\
$w_g(x,\, \mu)/w_0(x, \, \mu)$ has a pole at $x=x_M$ and the ratio
$w_\pi(x,\, \mu)/w_0(x, \, \mu)$ has a double pole
at the upper endpoint, which in particular
reflects the difference in the kinematics of decay of a free heavy
quark and of a heavy quark bound in hadron$^{\cite{bsvu,bsuv}}$. This
implies that the spectrum close to the endpoint is sensitive to the
infrared hadron dynamics, while in integral quantities, like the moments
of the lepton spectrum, the effects of this dynamics are integrated over
and are present only in the form of small corrections. Naturally, this
conclusion is valid only if the number $n$ of the moment is not
parametrically large, since high moments measure the spectrum near the
upper endpoint, and all the infrared effects come back. In the
expressions for the moments this growth of sensitivity to large
distances reveals itself in the growth with $n$ of the relative
magnitude of the non-perturbative corrections. Therefore, one can
consider as `safe' the moments with such $n$, for which the
non-perturbative correction is still small. As will be discussed, if one
chooses to keep individual terms in the corrections at a level below
10\% - 15\%, this would limit the range of $n$ to $n \le 5$. Thus
in what follows explicit results are presented for the moments in this
range of $n$.

According to eq.(\ref{g1}) ratio of the $n$-th moment to $M_0$
(the total rate) $r_n=M_n/M_0$ can be written as
\beq
r_n=r_n^{(0)} \left ( 1- {2 \over 3} {\alpha_s \over
\pi}\, \delta^{(1)}_n + {\mu_\pi^2 \over m_b^2} \, \delta^{(\pi)}_n+
{\mu_g^2 \over m_b^2} \, \delta^{(g)}_n \right)~,
\label{r1}
\eeq
where $r_n^{(0)}$ is the same ratio in the lowest approximation:
\beq
r_n^{(0)}=\left ( {m_b \over 2} \right )^n { {\int_0^{1-\mu^2}
 w_0(x,\,\mu)\, x^n \, dx} \over {\int_0^{1-\mu^2} w_0(x,\,
\mu)\, dx}}
\label{r0}
\eeq
and the corrections $\delta^{(1)}_n$, $\delta^{(\pi)}_n$ and
$\delta^{(g)}_n$ each being a function of $\mu$
are obtained from integrals with the corresponding
correction function $w(x,\,\mu)$ in eq.(\ref{g1}) as
\beq
\delta_n={{\int_0^{1-\mu^2} w(x,\,\mu)\, x^n \, dx} \over
{\int_0^{1-\mu^2} w_0(x,\,\mu)\, x^n \, dx}} -
{{\int_0^{1-\mu^2} w(x,\,\mu) \, dx} \over
{\int_0^{1-\mu^2} w_0(x,\,\mu) \, dx}}~.
\label{deln}
\eeq

The moments of the lowest order function $w_0(x,\, \mu)$ for $n \le 5$
are listed in the Appendix. The correction coefficients
$\delta^{(\pi)}_n$ can in fact be found in a simple analytical form.
This is possible due to the fact that the function $w_\pi (x, \, \mu)$
is related to a modification  by a small boost with $<{\bf v}^2> =
\mu_\pi^2/m_b^2$ of the lepton spectrum described by the function
$w_0(x, \, \mu)\,^{\cite{bsvu,bsuv}}$:
\beq
w_1(x,\, \mu)={x^2 \over 2} \, { \partial \over {\partial x}} \left (
{w_0(x,\, \mu) \over x } \right ) + {x^3 \over 6} \,
{ \partial^2 \over {\partial x^2}} \left (
{w_0(x,\, \mu) \over x } \right ) - {{w_0(x,\, \mu) }\over 2 }~.
\label{w1}
\eeq
Integrating by parts one readily finds that $\delta^{(\pi)}_n$
does not depend on $\mu$ and is given by
\beq
\delta^{(\pi)}_n=n(n+2)/6~.
\label{delpi}
\eeq

The expression for the coefficients $\delta^{(g)}_n$ can be found in a
somewhat lengthy analytical form.  The integrals in eq.(\ref{deln})
with the function $w_g (x,\, \mu)$ and $n \le 5$ are listed in the
Appendix. Similar integrals for
the perturbative coefficients $\delta^{(1)}_n$ can also, perhaps, be
done analytically as a function of the mass ratio $\mu$. However,
judging by the expression$^{\cite{jk,cj}}$ for the
function $w_1(x,\,\mu)$ and
by the analytical expression for the dependence of the $O(\alpha_s)$
correction to the total rate$^{\cite{nir}}$, the resulting formulas
should be prohibitively lengthy. For the practical purpose of analyzing
experimental data it is sufficient however to have a table of
these coefficients for values of $\mu$ around the approximate actual
value $\mu \approx 0.3$.
The numerical values of the coefficients $\delta^{(1)}_n$
and $\delta^{(g)}_n$ for $n \le 5$ are given in Tables 1 and 2.
Since for each $n$ these coefficients are
slowly varying functions of $\mu$, the tabular values can be used for an
interpolation.

One can see from these numerical values and from eq.(\ref{delpi}) that
for $\alpha_s \approx 0.2$, $\mu_g^2/m_b^2 \approx 0.015$ and
$\mu_\pi^2/m_b^2
\approx 0.015\, - \, 0.025$ the perturbative correction to the ratia $r_n$
is quite small as compared to the non-perturbative terms, and that each
of the latter terms is within 10\% - 15\% range for $n=5$, though the
overall non-perturbative correction is significantly smaller due to a
partial cancellation between the two terms.

\section{Discussion}

The estimates presented above illustrate that both the perturbative and
the non-perturbative QCD corrections are sufficiently small and
controllable in a number of ratia of moments of the lepton spectrum in
semileptonic $B$ decays, which number is sufficient for a detailed
experimental
study of the kinematical parameters of these decays. A simple numerical
inspection reveals that the ratia $r_n$ are in fact sensitive to the
quark mass difference $m_b-m_c$ rather than to the individual quark
masses. Therefore it is quite likely, that the value of this mass
difference can be determined with high precision from experimental data,
while to separate each of the masses, one will have to rely on other
types of analyses, e.g. on the existing determination of $m_b$ from the
$\Upsilon$ sum rules, or, possibly, on one from the inclusive spectrum
of photons, generated by the process $b \to s\, \gamma$, which may
become possible in a future development of the experiment \cite{cleo}.

One last remark is in order concerning a possible experimental
measurement of the moments $M_n$ in $e^+e^-$ annihilation at the
$\Upsilon(4S)$ resonance. Since the resonance is slightly above the
$B \overline B$ threshold, the $B$ mesons have momentum of about $0.3\,
GeV$, and their measured lepton spectrum is slightly distorted by boost.
However in order to account for this boost in the integral quantities
like the moments $M_n$ there is no need to transform the measured lepton
energy distribution to the $B$ rest frame. The reason is that a small
boost the expressions for the moments in the laboratory frame remain
valid after adding in quadrature the `intrinsic' average momentum
squared of the $b$ quark in meson, $\mu_\pi^2$ with that of the $B$
meson in the laboratory frame $ <{\bf p}^2>$. This obviously amounts to
replacing the $\mu_\pi^2$ by the effective quantity
\beq
{\overline \mu_\pi^2} = \mu_\pi^2+<{\bf p}^2>~.
\label{muef}
\eeq
One can notice that at the energy of the $\Upsilon(4S)$ resonance
the effect of the boost: $<{\bf p}^2> \approx 0.09
GeV^2$, is rather small in comparison with $\mu_\pi^2$.

\section{Acknowledgments}
I am thankful to M. Shifman, A. Vainshtein and N. Uraltsev for
discussions of heavy quark decays and to M. Danilov, G. Kostina and Yu.
Zaitsev for discussions of experimental possibilities.
This work is supported, in part, by the DOE grant DE-AC02-83ER40105.

\appendix
\section{Appendix}
The moments of the lowest-order energy distribution function,
$I^{(0)}_n=\int_0^{1-\mu^2} w_0(x,\, \mu) \, x^n \, dx$, entering
eq.(\ref{r0}), for $n \le 5$ are given by the following expressions
\beq
I^{(0)}_0=1 - 8\,{\mu^2} + 8\,{\mu^6} - {\mu^8} -
12\,{\mu^4}\,\log ({\mu^2})
\eeq
\beq
I^{(0)}_1=
{7\over {10}} - {{15\,{\mu^2}}\over 2} - 12\,{\mu^4} + 20\,{\mu^6} -
  {{3\,{\mu^8}}\over 2} + {{3\,{\mu^{10}}}\over {10}} - 6 \,\mu^4 \,
  \left( 3+\mu^2 \right) \,\log ({\mu^2})
\eeq
\beq
I^{(0)}_2=
{8\over {15}} - {{36\,{\mu^2}}\over 5} - 26\,{\mu^4} + 32\,{\mu^6} +
  {{4\,{\mu^{10}}}\over 5} - {{2\,{\mu^{12}}}\over {15}} -8\, \mu^4 \,
  \left( 3 +2\,{\mu^2} \right) \,\log ({\mu^2})
\eeq
\beq
I^{(0)}_3=
{3\over 7} - 7\,{\mu^2} - {{83\,{\mu^4}}\over 2} +
  {{85\,{\mu^6}}\over 2} + 5\,{\mu^8} + {\mu^{10}} -
{{{\mu^{12}}}\over 2} +
  {{{\mu^{14}}}\over {14}} -30\, \mu^4 \,
\left( 1+\mu^2 \right) \,
   \log ({\mu^2})
\eeq
\beq
I^{(0)}_4=
{5\over {14}} - {{48\,{\mu^2}}\over 7} - {{291\,{\mu^4}}\over 5} +
  {{252\,{\mu^6}}\over 5} + 15\,{\mu^8} - {\mu^{12}} +
  {{12\,{\mu^{14}}}\over {35}} -
{{3\,{\mu^{16}}}\over {70}} -12\, \mu^4 \,
  \left( 3 +4\,{\mu^2} \right) \,\log ({\mu^2})
\eeq
\begin{eqnarray}
I^{(0)}_5=
 {{11}\over {36}} - {{27\,{\mu^2}}\over 4} - {{759\,{\mu^4}}\over {10}} +
  {{329\,{\mu^6}}\over 6} + {{63\,{\mu^8}}\over 2} -
  {{7\,{\mu^{10}}}\over 2} - {{7\,{\mu^{12}}}\over 6} +
  {{9\,{\mu^{14}}}\over {10}} - {{{\mu^{16}}}\over 4} +
  {{{\mu^{18}}}\over {36}} - \nonumber \\
14 \,\mu^4 \,
 \left( 3+5\,{\mu^2} \right) \,
   \log ({\mu^2})~.~~~~~~~~~~~~~~~~~~~~~~~~~~~~~~~~~~
\end{eqnarray}

The first moments of the function $w_g$: $I^{(g)}_n=\int_0^{1-\mu^2}
w_g(x,\, \mu)\, x^n \, dx$, necessary for calculation of the
coefficients $\delta^{(g)}_n$, are given by the following expressions
\beq
I^{(g)}_0=
-{3\over 2} + 4\,{\mu^2} - 12\,{\mu^4} + 12\,{\mu^6} -
  {{5\,{\mu^8}}\over 2} - 6\,{\mu^4}\,\log ({\mu^2})
\eeq
\beq
I^{(g)}_1=
-2 + {{5\,{\mu^2}}\over 3} - 4\,{\mu^4} + 8\,{\mu^6} -
  {{14\,{\mu^8}}\over 3} + {\mu^{10}} - 4\,{\mu^2}\,\log ({\mu^2})
\eeq
\beq
I^{(g)}_2=
-{{104}\over {45}} - {{8\,{\mu^2}}\over 3} + 25\,{\mu^4} -
  {{40\,{\mu^6}}\over 3} - {{28\,{\mu^8}}\over 3} +
  {{16\,{\mu^{10}}}\over 5} - {{5\,{\mu^{12}}}\over 9} -4\,\mu^2 \,
  \left( 2 - 3\,{\mu^2} - {{10\,{\mu^4}}\over 3} \right) \,
   \log ({\mu^2})
\eeq
\begin{eqnarray}
I^{(g)}_3=
-{{53}\over {21}} - {{42\,{\mu^2}}\over 5} + {{155\,{\mu^4}}\over 2} -
  {{95\,{\mu^6}}\over 2} - 25\,{\mu^8} + 8\,{\mu^{10}} -
  {{73\,{\mu^{12}}}\over {30}} + {{5\,{\mu^{14}}}\over {14}} -
\nonumber \\
2 \, \mu^2 \,
  \left( 6 - 15\,{\mu^2} - 25\,{\mu^4} \right) \,\log ({\mu^2})
{}~~~~~~~~~~~~~~~~~~~~~~~~~~~~~~~
\end{eqnarray}
\begin{eqnarray}
I^{(g)}_4=
-{{75}\over {28}} - {{76\,{\mu^2}}\over 5} +
  {{1553\,{\mu^4}}\over {10}} - 86\,{\mu^6} - {{395\,{\mu^8}}\over 6} +
  20\,{\mu^{10}} - {{73\,{\mu^{12}}}\over {10}} +
  {{206\,{\mu^{14}}}\over {105}} - {{{\mu^{16}}}\over 4} -
\nonumber \\
2 \, \mu^2 \,
  \left( 8 - 27\,{\mu^2} - 60\,{\mu^4} \right) \,\log ({\mu^2})
{}~~~~~~~~~~~~~~~~~~~~~~~~~~~~~~~~
\end{eqnarray}
\begin{eqnarray}
I^{(g)}_5=
-{{151}\over {54}} - {{160\,{\mu^2}}\over 7} +
  {{1299\,{\mu^4}}\over 5} - {{1057\,{\mu^6}}\over 9} -
  {{455\,{\mu^8}}\over 3} + 49\,{\mu^{10}} - {{175\,{\mu^{12}}}\over 9} +
\nonumber \\
  {{103\,{\mu^{14}}}\over {15}} - {{23\,{\mu^{16}}}\over {14}} +
  {{5\,{\mu^{18}}}\over {27}}
-4\, \mu^2 \,
  \left( 5 - 21\,{\mu^2} - {{175\,{\mu^6}}\over 3} \right) \,
   \log ({\mu^2})~.
\end{eqnarray}

\newpage
\begin{table}
\begin{center}
\begin{tabular}{|c|ccccc|}
\hline
$m_c/m_b$ & $\delta^{(1)}_1$ & $\delta^{(1)}_2$ & $\delta^{(1)}_3$ &
$\delta^{(1)}_4$ & $\delta^{(1)}_5$ \\
\hline
0.25 & 0.0252 &  0.0669 &  0.1193 &  0.1787 &  0.2427 \\
0.26 & 0.0226 &  0.0619 &  0.1120 &  0.1693 &  0.2312 \\
0.27 & 0.0201 &  0.0571 &  0.1052 &  0.1603 &  0.2202 \\
0.28 & 0.0178 &  0.0527 &  0.0987 &  0.1519 &  0.2098 \\
0.29 & 0.0156 &  0.0485 &  0.0925 &  0.1438 &  0.1999 \\
0.30 & 0.0135 &  0.0446 &  0.0867 &  0.1362 &  0.1904 \\
0.31 & 0.0116 &  0.0408 &  0.0812 &  0.1289 &  0.1813 \\
0.32 & 0.0099 &  0.0373 &  0.0760 &  0.1219 &  0.1726 \\
0.33 & 0.0082 &  0.0340 &  0.0711 &  0.1153 &  0.1643 \\
0.34 & 0.0066 &  0.0309 &  0.0663 &  0.1090 &  0.1563 \\
0.35 & 0.0052 &  0.0280 &  0.0619 &  0.1029 &  0.1486 \\
\hline
\end{tabular}
\end{center}
\caption{Numerical values of the perturbative correction coefficients
$\delta^{(1)}_n$ in eq.(11) for $n \le 5$ and $\mu=m_c/m_b$ in the
range from 0.25 to 0.35\,.}
\end{table}

\begin{table}
\begin{center}
\begin{tabular}{|c|ccccc|}
\hline
$m_c/m_b$ & $\delta^{(g)}_1$ & $\delta^{(g)}_2$ & $\delta^{(g)}_3$ &
$\delta^{(g)}_4$ & $\delta^{(g)}_5$ \\
\hline
0.25 & -1.188  & -2.419 &  -3.674 &  -4.94  &  -6.213 \\
0.26 & -1.187  & -2.414 &  -3.664 &  -4.926 &  -6.195 \\
0.27 & -1.186  & -2.411 &  -3.657 &  -4.915 &  -6.17  \\
0.28 & -1.185  & -2.408 &  -3.651 &  -4.905 &  -6.165 \\
0.29 & -1.185  & -2.406 &  -3.647 &  -4.898 &  -6.155 \\
0.30 & -1.185  & -2.405 &  -3.644 &  -4.893 &  -6.147 \\
0.31 & -1.186  & -2.406 &  -3.643 &  -4.89  &  -6.142 \\
0.32 & -1.187  & -2.407 &  -3.644 &  -4.89  &  -6.14  \\
0.33 & -1.189  & -2.409 &  -3.646 &  -4.891 &  -6.141 \\
0.34 & -1.192  & -2.413 &  -3.65  &  -4.895 &  -6.144 \\
0.35 &-1.195   & -2.417 &  -3.655 &  -4.901 &  -6.151 \\
\hline
\end{tabular}
\end{center}
\caption{Numerical values of the non-perturbative correction
coefficients
$\delta^{(g)}_n$ in eq.(11) for $n \le 5$ and $\mu=m_c/m_b$ in the
range from 0.25 to 0.35\,.}
\end{table}
\end{document}